\documentclass[11pt]{article}
\usepackage{graphicx}
\usepackage{overcite}
\newcommand{\A}{\alpha}
\newcommand{\ab}[1]{\mathrm{#1}}
\newcommand{\Td}{T_{\mathrm{diff}}}              
\newcommand{\Tr}{T_{\mathrm{reac}}}
\newcommand{\Ne}{N_{\mathrm{e}}}
\newcommand{\Nc}{N_{\mathrm{c}}}   
\newcommand{\T}{\tau_0}
\newcommand{\ts}{\tilde{\sigma}}
\newcommand{\tn}{\tilde{n}}
\newcommand{\Rb}{R_{\mathrm{b}}}    

\begin{document}
\title{\textsf{A Scaling Theory of the Competition between Interdiffusion and Cross-Linking at Polymer Interfaces}}
\author{A.\ Aradian\footnote{Achod.Aradian@college-de-france.fr}, E.\ Rapha\"{e}l
and P.-G.\ de Gennes\\ Physique de la Mati\`ere Condens\'ee, 
C.N.R.S.\ URA 792, \\ Coll\`{e}ge de France, 11 place Marcelin 
Berthelot, \\ 75231 Paris Cedex 05, France} \maketitle 
\begin{abstract}
We study theoretically situations where competition arises between 
an interdiffusion process and a cross-linking chemical reaction at 
interfaces between pieces of the same polymer material. An example 
of such a situation is observable in the formation of latex films, 
where, in the presence of a cross-linking additive, colloidal 
polymer particles initially in suspension come at contact as the 
solvent evaporates, and, optimally, coalesce into a continuous 
coating. We considered the low cross-link density situation in a 
previous paper (Aradian, A.; Rapha\"{e}l, E.; de Gennes, P.-G. 
{\it Macromolecules} \textbf{2000}, {\it 33}, 9444), and presented 
a simple control parameter that determines the final state of the 
interface. In the present article, with the help of simple scaling 
arguments, we extend our description to higher cross-link 
densities. We provide predictions for the strength of the 
interface in different favorable and unfavorable regimes, and 
discuss how it can be optimized. 
\end{abstract}
\section{Introduction}
The purpose of this article is to present a theoretical approach 
of situations where both interdiffusion and chemical cross-linking 
occur at the interface between two polymer pieces put in contact 
at a temperature above their glass transition. Such situations are 
rather commonplace in polymer processing. More precisely, our main 
motivation lies in the formation of latex coatings:~\cite 
{WinnikReview} these contain many such interfaces (of mesoscopic 
dimensions), and the issue of simultaneous interdiffusion and 
cross-linking between adjacent polymer particles is crucial for 
the final properties of the product. As shall be seen, our 
theoretical description remains rather generic, and should thus, 
in addition to latex coatings, be of some relevance in other 
related contexts, like the formation of a (macroscopic) joint 
between two pieces of rubber. 

The formation steps of latex coatings can be summarized as 
follows~\cite{WinnikReview} (see Figure 1). A colloidal dispersion 
of polymer particles in water is applied onto a substrate. As the 
water evaporates, the particles come into contact, and deform to 
create a void-free array of polyhedral cells. When good contacts 
are formed, neighboring particles start to coalesce, i.e.\ chain 
interdiffusion takes place at the microscopic 
scale,~\cite{JonesRichards, 
Woolbook,PGGCras1980,PGGSanchez,Prager,Wool} until, finally, the 
initial interfaces have ``healed'' completely.
    \begin{figure}
    \centering
    \includegraphics*[width=12cm, clip=true, bb=3.5cm 11cm 18cm 
    19cm]{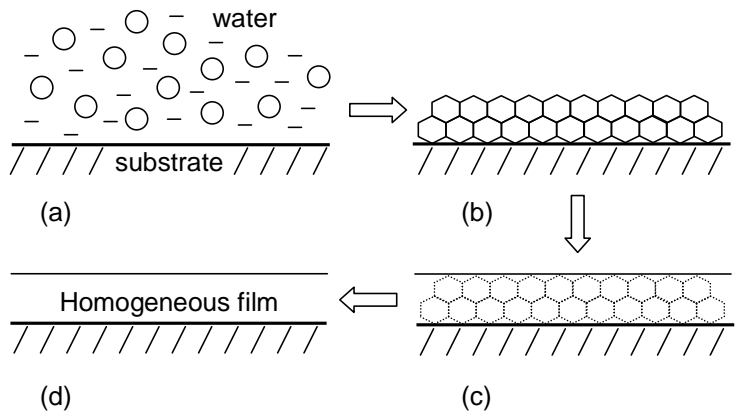} 
    \caption{Formation of a latex coating (in the absence of 
an external cross-linker agent). (a)~The initial colloidal 
dispersion. (b)~The particles form a void-free array upon solvent 
evaporation. (c)~Neighbouring particles coalesce, and the initial 
boundaries faint gradually. (d)~Ultimately, the initial granular 
structure has been lost, and the film is continuous.} 
    \end{figure} 
It is also quite usual to add a cross-linker agent into the system 
(directly in the initial dispersed 
state),~\cite{Bufkin,TaylorWinnik} with the main goal of improving 
the bulk properties of the material. However, the introduction of 
a cross-linker brings along difficulties: at the interfaces 
between neighboring cells, a competition will arise between the 
interdiffusion process and the chemical cross-linking reaction, as 
they will both proceed in parallel. The reason for this fact is 
simple: since the mobility of polymer chains in the entangled 
state is drastically reduced by chain branching, the cross-linking 
slows interdiffusion down. Therefore, the addition of cross-linker 
into the formulation, by preventing the healing of the interfaces, 
and even any coalescence at all, may sometimes prove more harmful 
than beneficial. The central issue in this respect is to control 
the timing of the chemical reaction. 

In a previous paper,~\cite{Aradian} we addressed this issue and 
proposed a control parameter $\A$, related to the physico-chemical 
properties of the polymer and the cross-linker. We also discussed 
two limiting regimes, namely the favorable ``slow-reaction'' 
regime and the unfavorable ``fast-reaction'' regime, and computed 
the final state of coalescence in the film in each of them. Our 
study had however a strong limitation: it assumed a very low 
cross-linker concentration in the system, of the order of 
\emph{one} cross-linker molecule per chain. Obviously, practical 
situations are much more varied --- in applications, there is 
often a much higher cross-link density in the final material. The 
present article is aimed at extending our description by including 
this feature. In order to clarify as much as possible the physical 
content of the theory, we will develop an approach based mainly on 
scaling arguments. 
                                                                                     
Our point of view in this article will be that, when a particular 
application is intended for the material under consideration, one 
is led to specify requirements on the \emph{bulk} properties of 
the material. Now, in a quite general fashion, these bulk 
properties are mainly controlled by the final density of 
cross-links present in the material. As a consequence, the 
specifications on the bulk properties impose, in turn, that a 
given density of cross-links must be incorporated in the material. 
In what follows, we will thus work with the assumption that the 
cross-link density has been fixed to a given value beforehand (for 
``external'' reasons) and is a given parameter of the problem. 
More precisely, we will use the number of monomers between 
cross-links $\Nc$ (inversely proportional to the density) as a 
given parameter. We will then provide estimates of the interfacial 
energy $G$ under this constraint on $\Nc$, and see how it can be 
optimized. In other words, we could say that we try to optimize 
the interface strength (and all related properties of the 
material) at given bulk characteristics. 

In order to explain this approach of the problem on a more 
concrete basis, we can invoke again the context of latex coatings. 
Examples of ``bulk related'' properties are given by the Young 
modulus of the coating, the surface hardness, or the extent of 
swelling in the presence of solvent.~\cite{Brown} Experimentally, 
these properties are mainly determined by the density of 
cross-links in the material, and are enhanced as this density 
increases.~\cite{Brown} For the Young modulus $E$, the relation is 
in fact well-known from the classical theory of rubber elasticity: 
$E \sim kT/(\Nc a^3)$, where $kT$ is the thermal energy, $a$ the 
polymer unit (monomer) size and $\Nc$ the number of units between 
cross-links. On the other hand, there is another set of 
properties, like the film thoughness~\cite{Zosel}, the tensile 
strength~\cite{Kim}, the resistance to scratching and solvent 
application,~\cite{Brown} or the film homogeneity and aspect, that 
are rather related to the state of the interfaces. Such 
``interface related'' properties are strongly dependent on the 
extent of coalescence inside the film, or, equivalently, on the 
strength of the interfaces (the stronger the interfaces, the 
better these properties). 

We will assume in the rest of this paper that the coalescence and 
interface strength, which determine the level of the interface 
properties, are fairly well reflected by the value of the 
\emph{interfacial adhesion energy} $G$ between two neighboring 
particles of the coating. Predictions on this interfacial energy 
in different regimes constitute the main goal of this work. 

The article is organized as follows. In section~2, we derive for a 
given distance between cross-links $\Nc$, and with the help of 
very simple scaling arguments, the control parameter $\A$ that 
determines the final state of the film. We also compute important 
quantities like the interpenetration length and the surface 
density of crossing chains. In section~3, we estimate the 
interfacial energy $G$ in different regimes. Finally, in 
section~4, we discuss our results and present further perspectives 
of study. 
\section{A scaling approach to the interdiffusion/cross-linking competition}
In this section, we present simple scaling arguments describing 
the interdiffusion/cross-linking competition at polymer-polymer 
interfaces. Let us first describe more precisely the physical 
system under consideration. 
\subsection{Description of the system}
In the context of latex coatings, several kind of systems are 
possible: mixtures of particles of the same or of different 
polymers (either miscible or immiscible), reaction through 
functionalized groups on the chains or by addition of a separate 
cross-linker molecule, etc. As in our previous study, we will here 
remain with one of the simplest systems of experimental relevance, 
that is to say, homogeneous particles of the same polymer, and 
cross-linking by addition of an external chemical reagent. (A 
brief discussion of other systems is given in section 4.3 of 
ref.~\citen{Aradian}.) In the system chosen, all interfaces are 
identical, and we focus on the evolution of one of them --- which 
amounts to studying the general problem of the competition of 
interdiffusion and ``external'' cross-linking at a symmetric 
polymer/polymer interface. 

Let us thus assume that at $t = 0$, two pieces of the same polymer 
material (for instance, two particles from a latex dispersion) are 
put into contact at a temperature above the glass transition, 
forming an interface that will be taken to be planar. The polymer 
pieces are made of a fixed initial network, plus a volume 
concentration $\rho_0$ of free chains (which will be able to 
diffuse across the interface). The free chains are assumed to be 
linear and monodisperse in the initial state (before the 
cross-linking reaction starts). In order to remain close to 
experimental situations, the chains are taken as statistical 
copolymers of two different units A and A*, of which only A* may 
bind with the cross-linker. As a consequence of this chain 
structure, a quantity that will naturally prove of importance is 
$A_0^*$, the initial volume concentration of A*-sites in the 
system. The total number of units per chain is $N$, with $N$ 
greater than the entanglement threshold $\Ne \simeq 100$ 
(entangled state). 

The cross-linker agent, X, is bi-functional, and we may split the 
cross-linking reaction in two stages. The first one is the binding 
of an X molecule to an A*-site borne by a polymer chain P$_1$, and 
can be written as P$_1$A*+X $\to$ P$_1$A*--X. The second step is 
the effective cross-linking reaction, between a polymer chain 
P$_1$ bearing an A*X-site, and an A* on another chain P$_2$, i.e. 
P$_1$A*--X + A*P$_2$ $\to$ P$_1$A*--X--A*P$_2$. The first step is 
quite a fast one, as it involves the diffusion of small X 
molecules. The second step, a reaction between two macromolecules, 
will on the contrary be the limiting one. Thus, in the following, 
we will consider the kinetics of the second stage only, and take 
for granted that at $t = 0$, the first step is completed. As 
explained already, the number of monomers between cross-links 
$\Nc$ (reached when the cross-linking reaction is completed) is 
fixed, and, we naturally suppose $\Nc \leq N$ (all chains are 
branched and form a network at the end). 

Finally, we consider free, unreacted, chains to be mobile, in 
contrast with chains that have been subject to branching at some 
time (due to the cross-linking reaction), which are from that time 
on considered to be \emph{fixed} at the location of occurrence of 
the reaction. Such a simplification is not unreasonable, since it 
is known that the reptation motion of a branched object is 
exponentially slower than that of a linear one.~\cite{PGGStar} 
\subsection{The control parameter $\A$}

Following the tracks of our previous work,~\cite{Aradian} we here 
derive a parameter $\A$, called the control parameter, which 
characterizes the final state of the interface (once all chains 
have reacted and are immobilized). The idea is rather 
straightforward: the important parameter of the problem is given 
by the comparison of the rate of the interdiffusion with the rate 
of the cross-linking reaction. Thus, we define $\A$ as the ratio 
between the typical interdiffusion time $\Td$ and the typical 
reaction time $\Tr$, that is 
    \begin{equation}
    \label{alphadef}
    \A = \frac{\Td}{\Tr}
    \end{equation}
There are obviously two limits: when the reaction is much faster 
than the interdiffusion (``fast-reaction'' regime), the chains are 
locked in place before any significant coalescence can occur. When 
the reaction is slow enough (``slow-reaction'' regime), the 
interface heals completely before the reaction freezes the system.
 
The interdiffusion time $\Td$ is the time needed to heal 
completely the initial interface between the pieces in contact, in 
the absence of reaction. Such a healing occurs when interfacial 
chains on either side have traveled a distance comparable to their 
own size inside the neighboring piece of 
polymer.~\cite{JonesRichards,Woolbook,PGGCras1980,PGGSanchez,Prager,Wool} 
Referring to the theory of reptation,~\cite{PGGreptation, 
DoiEdwards} the time needed for a chain to travel over its own 
size is the so-called reptation time $T_{\ab{rep}}= \T N^3/\Ne$, 
(where $\T$ is a microscopic time typical of molecular agitation). 
Thus we deduce that
    \begin{equation}
    \label{Tdiff}
    \Td \simeq T_{\ab{rep}} = \T N^3/\Ne
    \end{equation}

We now need to estimate the typical reaction time $\Tr$. Quite 
intuitively, we define this time as the time required to observe 
\emph{one reaction per chain} in the system. From a scaling law 
point of view, after this typical time, every chain in the system 
has undergone branching at least once and is fixed in position: 
the final state of the film is reached. To evaluate $\Tr$, let us 
follow one A*X group on a given chain, chosen at random, and think 
in terms of a lattice model. Every $\T$ time intervals, this A*X 
group jumps onto a new site on the lattice, neighboring the one it 
occupied during the previous time interval. From this new 
location, the A*X is in position to react with all the A* lying 
within a ``capture distance'' $b$, which are, roughly, $A_0^* b^3$ 
in number. However, only a fraction of these ``collisions'' inside 
the capture sphere are efficient in giving a true reaction: this 
is reflected in the value of $Q$, defined as the reactivity of the 
cross-linker, which gives the probability of effective reaction 
between two partners per unit time spent in collision (that is, as 
long as their distance is less than the capture radius). Then, as 
between two jumps the A*X sojourns a time $\T$ on each location, 
the actual number of reactions at each location is $Q \T A_0^* 
b^3$ (for one A*X group). Remembering that each chain contains a 
number $N/\Nc$ of A*X groups, we deduce that the waiting time 
$\Tr$ necessary to reach a number of \emph{one} reaction \emph{per 
chain} is given by the relation $(\Nc/N) Q \T A_0^* b^3 (\Tr / \T) 
= 1$. From that, we obtain 
    \begin{equation}               
    \label{Treac}
    \Tr \simeq \frac{\Nc}{N} \frac{1}{Q A_0^* b^3}
    \end{equation}
The preceding expression shows that the chemical reaction is all 
the faster as the cross-linker is concentrated, the chains are 
long, the cross-linker is reactive and the concentration in sites 
capable of binding to the cross-linker is high.

The above estimates of the characteristic times finally yield the 
expression of the control parameter $\A$: 
    \begin{equation}
    \label{alpha} 
    \A = \frac{\Td}{\Tr} \simeq Q \T A_0^* b^3 \frac{N^4}{\Ne \Nc} 
    \end{equation} 
which will serve as a basis for the rest of the article.

For a given system, the two times $\Td$ and $\Tr$ should normally 
be experimentally measurable (and thus the numerical value of $\A$ 
could be known in practice). The diffusion time would be easy to 
determine, as we know from eq.~\ref{Tdiff} that $\Td$ is close to 
$T_\ab{rep}$. As for the reaction time $\Tr$, the unknown 
quantities are the microscopic reactivity $Q$ and the capture 
radius $b$. These can be obtained by monitoring the rate of the 
cross-linking reaction A*X + A* $\to$ A*XA*, and by extracting out 
the value of the corresponding reaction constant, 
$k$.~\cite{reactivity} One can then make use of the 
physicochemical relation~\cite{O'Shaughnessy} $k=Qb^3$ to compute 
directly the value of $\Tr$ from eq.~\ref{Treac} (the other 
quantities appearing in the equation being easily measured by 
other means). 

We also note that this formula for $\A$ is the generalization of 
the one derived in the low cross-linker limit in 
ref.~\citen{Aradian} (which is retrieved by having $\Nc$ as large 
as allowed, i.e. $\Nc \simeq N$). 
\subsection{Interpenetration length and crossing chains density}
In section~3, we will compute interfacial adhesion energies that 
represent the strength of the interface(s). In the process, as 
explained there, we will need to know two quantities, which are 
the interpenetration length (the length over which the two polymer 
pieces forming the joint have mixed), and the crossing density 
(the density of chains crossing the interface per unit area), at 
the end of the interface evolution. We now give a scaling
derivation of these quantities. 

In a way that is consistent with the previous derivation of the 
control parameter $\A$, the scaling argument presented here 
simplifies the kinetics of the chemical reaction as follows: as 
long as the time $t$ is less than the reaction time $\Tr$, the 
reaction does not occur, and all initially free chains remain 
free. Then, the reaction occurs precisely at $t = \Tr$, and after 
that time, all chains are fixed, meaning that the final state of 
the film is reached. (More refined calculations validating this 
scaling approach can be found in ref.~\citen{Aradian}). 

Let us start with the interpenetration length, which is the 
average thickness that chains from one side of the interface 
achieve to invade inside the other side. We know from reptation 
theory~\cite{PGGreptation,DoiEdwards} that, in the entangled 
state, each chain undergoes a one-dimensional diffusion inside a 
contorted "tube" representing the topological constraints imposed 
by the other chains. Thus, in a time $t$, a chain travels a 
curvilinear distance $S(t) \sim t^{1/2}$ inside the tube. To this 
contorted distance corresponds a linear, ``as the crow flies'', 
distance $L(t)$, which can be shown to be proportional to the 
square root of $S(t)$ (because the tube shape is a random walk). 
With the appropriate physical constants, the interpenetration 
length $L(t)$ at time $t$ can be 
written~\cite{Woolbook,PGGCras1980,PGGSanchez,Prager,Wool} 
    \begin{equation}
    \label{distanceinterpenetration}
    L(t) \simeq \sqrt{a S(t)} \simeq R_0 \left( t/T_{\ab{rep}} 
    \right)^{1/4} \quad (\A \ll 1)
    \end{equation}
where $a$ is the monomer size, $R_0$ is the Gaussian size of the 
chains ($R_0 = a N^{1/2}$), and $T_{\ab{rep}}$ is the reptation 
time ($T_\ab{rep} \simeq \Td$, see eq.~\ref{Tdiff}). Thus, when 
the reaction at $t = \Tr$ brings an end to diffusion, the final 
interpenetration length is 
    \begin{equation}
    \label{finalinterpenetration}
    L_{\ab{final}} \simeq R_0 \left( \Tr/T_{\ab{rep}} \right)^{1/4} 
    \simeq \frac{R_0}{\A^{1/4}} \quad (\A \ll 1)
    \end{equation}
and the corresponding contorted distance, really traveled by the 
chain units along the tube, is 
    \begin{equation}
    \label{Sfinal}
    S_{\ab{final}} \simeq L_{\ab{final}}^2/a \quad (\A \ll 1)
    \end{equation} 

We keep these equations for further use, and now turn to the 
computation of the density $\sigma$ of the chains that have 
crossed the interface (per unit area of surface). To find 
themselves beyond the interface at time $t = \Tr$, these chains 
were all necessarily not farther than a distance $L_{\ab{final}}$ 
from the interface, at $t = 0$. Assuming that the initial volume 
concentration of free chains $\rho_0$ is uniformly distributed 
inside the material,~\cite{chainendsdistri} we deduce that the 
surface density of crossing chains is 
    \begin{equation}
    \label{crossingdensity}
    \sigma \simeq \rho_0 L_{\ab{final}} \quad (\A \ll 1)
    \end{equation}

To conclude this section, we should caution that, as indicated, 
the previous equations are actually valid (and will be used) only 
for $\A \ll 1$, that is in the slow-reaction regime. When $\A \gg 
1$, a saturation occurs (for example, in the crossing density 
$\sigma$), because the rate to which polymer chains enter the 
interface becomes equal to the rate to which those already present 
there leave it.~\cite{Aradian} However, we do not need to enter 
into these details, as it is possible to compute the adhesion 
energy at $\A \gg 1$ without further knowledge. 
\section{Different regimes for the interfacial adhesion energy}
We are now in a position to compute the interfacial adhesion 
energy, which we consider as a fair indicator of the extent of 
coalescence and the strength developed at the interface. (We 
remind the reader that these quantities are important notably in 
the context of latex coatings, where they have a strong impact on 
many properties). 
\subsection{A brief reminder on scission and extraction}
The interfacial adhesion energy is the energy one has to provide 
to open a fracture at the interface between the two polymer pieces 
at contact (at the end of the interface evolution). Because of 
visco-elastic dissipation, this energy generally depends on the 
rate at which the fracture is propagated. We shall here focus on 
the quasi-static energy $G$, i.e.\ when the rate approaches zero. 
The energy cost in this limit has two contributions (in 
elastomers). One is the classical Dupr\'e's work, valid for all 
materials (including those made of short molecules), and due to 
van der Waals interactions. This is a constant offset term that we 
will henceforth omit. The other contribution is specific to 
polymers, and comes from the fact that some polymer chains 
(``connectors'') may cross the interface and extend over both 
sides, thereby strengthening the joint. This connector 
contribution itself finds an origin in two distinct processes 
taking place at the opening of the fracture: chain scission 
(rupture of chemical bonds) and chain extraction (connectors are 
dragged out of the surrounding matrix). At zero-detachment rate, 
whether a given interfacial chain will undergo pull-out or 
scission depends on whether it is chemically tethered, 
respectively, on one or both sides of the interface (Figure 2).
    \begin{figure}
    \centering
    \includegraphics*[clip=true, bb=6.5cm 12cm 14.5cm 
    17.5cm]{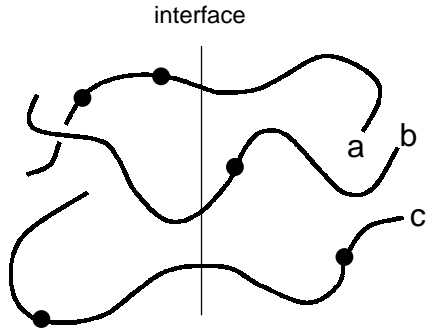} 
    \caption{Examples of connector chains. Chains ``a'' and ``b'' are 
tethered only on one side of the interface by chemical bonds 
(represented by black dots). At the fracture opening, they will be 
pulled out of the other side. Chain ``c'' is tethered on both 
sides, and thus will have to break (scission).} 
    \end{figure} 

Theoretical models are available for both chain scission and 
extraction. In the case of scission, Lake and 
Thomas~\cite{LakeThomas} found 
    \begin{equation}
    \label{scission}
    G_{\ab{scission}} \simeq  U_0 \ts \tn
    \end{equation}  
where $U_0$ is of the order of a typical chemical bond energy 
(denoted $U_\chi$), $\ts$ is the density of connectors per unit 
area of interface, and $\tn$ is the number of monomers under load 
along each connector. The physical origin of this expression is 
that, to rupture a chain between two cross-link points, 
\emph{each} monomer between these points must be put under load 
and brought close to the breakage threshold. Hence, to bring a 
chain to scission, we have to provide an energy $U_\chi$ to all 
the $\tn$ monomers between the anchorage points. For further use, 
we note that, in our case, the number of monomers under load is 
naturally fixed by the separation between cross-links, so that we 
have $\tn \simeq \Nc$ (for scission). 

We should also immediately point out to avoid later confusion, 
that the connector density $\ts$ is \emph{not} necessarily equal 
to the crossing density $\sigma$ given in 
eq.~\ref{crossingdensity}, which concerns polymer chains with $N$ 
units: for instance, a chain of $N$ units that crosses the 
interface brings (from the adhesion point of view) several 
connectors made of $\tn \simeq \Nc$ units. This distinction will 
be important to obtain a correct estimation of the adhesion energy 
in some cases. 

In the situation of chain extraction, Rapha\"{e}l and de 
Gennes~\cite{EliePGG} found a very similar formula: 
    \begin{equation}
    \label{pullout} 
    G_{\ab{extraction}} \simeq U_0 \ts \tn
    \end{equation}  
where $\ts$ is the density of connectors to be pulled out, $\tn$ 
the length to be extracted, but this time, the factor $U_0$ giving 
the energy scale is a van der Waals bond energy $U_{\ab{v}}$. 
Despite the analogy with eq.~\ref{scission}, the physics is quite 
different: when a chain is extracted, it is exposed to air 
(interfacial cost) and extended (entropy loss). At room 
temperature, both contributions to the energy are of the same 
order: $U_{\ab{v}}$ per monomer. 

In mixed situations where during the fracture process, some chains 
are extracted while others are broken, we will simply assume that 
the two types of dissipation are additive. However, in such 
instances, the adhesion energy $G$ is then often dominated by the 
chemical part, because $U_\chi/U_{\ab{v}} \simeq 100$. 
\subsection{Estimation of the interfacial energy $G$}
With the above models, we may now estimate the interfacial 
adhesion energy $G$ in different regimes. As announced earlier, we 
will distinguish two regimes depending on the value of the control 
parameter $\A$: the slow-reaction regime ($\A \ll 1$) and the 
fast-reaction regime ($\A \gg 1$). (It will also appear that the 
fast-reaction regime is itself divided in two sub-regimes.) We 
here present calculations, and defer a physical discussion of our 
findings to the next section. 

We start with the slow-reaction regime: the reaction is much 
slower than the interdiffusion process, which means that the 
interface has time to heal completely before the chains are 
stitched. In other words, the interface is allowed to reach the 
equilibrium state (relative to the interdiffusion process), and, 
ideally, it is no different from any other plane drawn inside the 
bulk of the material. Thus, it is natural that the adhesion energy 
of the interface becomes in this situation equal to the tear 
energy of the bulk (i.e., the tear energy of a network with a 
cross-link separation $\Nc$). In Figure~3, chain ``a'' illustrates 
a typical configuration of interfacial chains in this regime. When 
a fracture is opened, one will have to break (chemically) all 
these chains that cross the fracture plane, and we can use the 
Lake and Thomas formula (eq.~\ref{scission}), with a length of the 
connectors $\tn \simeq \Nc$, to estimate the adhesion energy. The 
chains that cross the interface (and undergo scission) lie inside 
a Gaussian radius $a \sqrt{\Nc}$ from the fracture, and as the 
number of chains of length $\Nc$ is $1/(\Nc a^3)$ per unit volume, 
we deduce that we have $\ts \simeq a \sqrt{\Nc} \times 1/(\Nc a^3) 
\simeq 1/(\sqrt{\Nc} a^2)$ broken connectors per unit area. Thus, 
the adhesion energy is simply (eq.~\ref{scission}) 
    \begin{figure}
    \centering
    \includegraphics*[clip=true, bb=6.5cm 11.5cm 15cm 
    17.5cm]{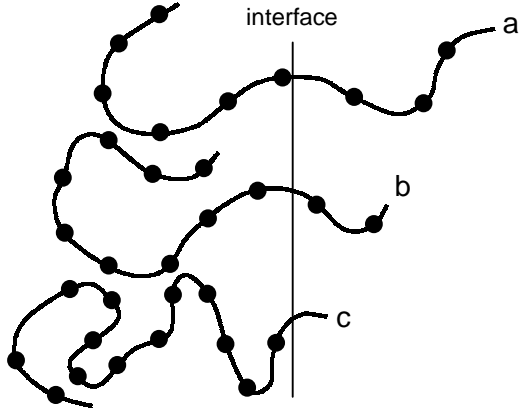} 
    \caption{Typical conformations of interfacial chains. Chain ``a'' 
is typical of the slow-reaction regime, with arms of comparable 
size on both sides of the interface. Chain ``b'' corresponds to 
the first sub-regime of the fast-reaction regime, with only a 
small portion of the chain inserted beyond the interface, and will 
have to rupture at the opening of a fracture. Chain ``c'' depicts 
the situation in the second sub-regime, where the inserted 
portions are even smaller. This chain will rather be extracted 
from beyond the interface.} 
    \end{figure} 
    \begin{equation}
    \label{Gslow}
    G \simeq \frac{U_0}{a^2} \sqrt{\Nc} 
    \simeq G_{\ab{max}} \qquad (\A < 1,\, N<N_1)
    \end{equation}
which is the highest possible value in the material (at fixed 
$\Nc$) and is independent of the value of the control parameter 
$\A$. We note that eq.~\ref{Gslow} refers to values of $\A 
< 1$. We may also say that, if all the parameters other than $N$ 
are fixed in the expression of $\A$ (eq.~\ref{alpha}), this 
corresponds equivalently to having $N$ less than a certain 
threshold $N_1$ (which we shall compute explicitly later on). 

Let us now consider the adhesion energy in the fast-reaction 
regime: here, the interface has been frozen in an 
out-of-equilibrium state. In average, chains did not have much 
time to diffuse across the boundary, and the inserted portions are 
shorter than in the previous slow-reaction regime (a sketch of a 
typical chain at the interface in the fast-reaction regime is 
provided by the chain labeled ``b'' in Figure~3). Again, there is 
scission, and the length of the connectors that will be broken at 
the fracture opening is still $\tn \simeq \Nc$. The density of 
chains crossing the interface is not at equilibrium, and has the 
value $\sigma \simeq \rho_0 L_{\ab{final}}$, as given by 
eq.~\ref{crossingdensity}. However, we should be careful at this 
point: as announced earlier, this crossing density $\sigma$ is 
\emph{not} the density of actual connectors $\ts$. The crossing 
density $\sigma$ represents the number of chains of length $N$ 
that cross the interface, but, once the cross-links are formed, 
each of these chains is actually divided into $N/\Nc$ segments of 
$\Nc$ monomers, and thus gives birth to \emph{several} connectors 
of length $\Nc$. To take this fact into account, it can be shown 
that the crossing density $\sigma$ must be multiplied by a factor 
equal to $L_{\ab{final}}/(a \sqrt{\Nc})$ to give the actual 
connector density $\ts$ (see the Appendix for a detailed 
calculation): in computing the adhesion energy, we must thus use 
$\ts \simeq L_{\ab{final}}/(a \sqrt{\Nc}) \times \sigma \simeq 
L_{\ab{final}}^2 \rho_0/(a \sqrt{\Nc})$. Estimating the initial 
volume density of free chains roughly as $\rho_0 \simeq 1/(N a^3)$ 
(within a numerical prefactor),~\cite{density} we find finally 
    \begin{equation}
    \label{Gfastscission}
    G \simeq U_0 \Nc \ts \simeq G_{\ab{max}} 
    \frac{1}{\A^{1/2}} \qquad (1<\A < \A_2,\, N_1<N<N_2)
    \end{equation}
We see that now the energy displays an inverse dependence in $\A$, 
and drops as $\A$ increases. 

However, as indicated, the previous equation is valid in a limited 
range, either in $\A$ $(\A < \A_2)$, or in $N$ $(N<N_2)$ (the 
critical values $\A_2$ and $N_2$ will be given shortly). The 
reason for this is that if the reaction becomes really very fast 
compared to the interdiffusion (as $\A$ or $N$ increases), the 
portions of interfacial chains that are allowed to penetrate into 
the other side of the interface are extremely short, and become 
shorter than $\Nc$. In such conditions, as exemplified by chain 
``c'' in Figure~3, these portions have almost no chance to be 
cross-linked on both sides of the boundary, and should be rather 
extracted at the opening of a fracture, implying that we must now 
consider a sub-regime of the fast-reaction regime where chain 
extraction predominates inside the adhesion energy 
(eq.~\ref{pullout}). The length $\tn$ to be pulled out is of the 
order of the \emph{curvilinear} length inserted beyond the 
interface, i.e. $\tn \simeq S_{\ab{final}}/a \simeq 
L_{\ab{final}}^2/a^2$ (eq.~\ref{Sfinal}). The density of 
connectors here has no complications: there is only \emph{one} 
chain end to be extracted per crossing chain, and so each crossing 
chains makes one connector. (Some chains may have inserted both 
ends beyond the interface, but this is probably unimportant.) The 
connector density is thus simply $\ts \simeq \rho_0 
L_{\ab{final}}$ (eq.~\ref{crossingdensity}), and altogether, the 
energy writes, using eq.~\ref{pullout}: 
    \begin{equation}
    \label{Gfastextraction}
    G \simeq U_0 \frac{L^2}{a^2} \ts \simeq \frac{U_0}{a^2} N^{1/2} 
    \frac{1}{\A^{3/4}} \quad (\A > \A_2, \, N>N_2)
    \end{equation} 
The inverse dependence of $G$ in $\A$ in this sub-regime becomes 
more pronounced. 

We now have, with 
eqs.~\ref{Gslow},~\ref{Gfastscission}~and~\ref{Gfastextraction}, a 
complete estimation of the interfacial adhesion energies in the 
whole range of the $\A$ parameter. The critical values of $\A$ 
marking the transitions between the various regimes are: 
$\A=\A_1=1$ where we cross-over from the slow- to the 
fast-reaction regime, and $\A = \A_2$ where, inside the 
fast-reaction regime, we switch from a scission-dominated 
sub-regime to an extraction-dominated one. 

The value of $\A_2$ is easy to determine, since the corresponding 
transition between sub-regimes is characterized by the fact that 
the number of monomers inserted beyond the interface by 
interfacial chains is equal to the number of monomers between 
cross-links, i.e. $L_{\ab{final}}^2/a^2 = \Nc$. Since 
$L_{\ab{final}} \simeq R_0/\A^{1/4}$ 
(eq.~\ref{finalinterpenetration}), we deduce that 
    \begin{equation}
    \label{alpha2}
    \A_2 = \left( \frac{N}{\Nc} \right)^2
    \end{equation}

Alternately, we can also think of these critical $\A$ values as 
critical values on the chain length $N$, if within our given 
system, the chain length is the only free parameter (all the other 
parameters entering the expression of $\A$ given in 
eq.~\ref{alpha} are fixed with known values). The cross-over from 
slow- to fast-reaction regime sets a first critical length $N_1$, 
which is determined by writing that $\A = 1$ explicitly in terms 
of all the parameters of the system (eq.~\ref{alpha}):
    \begin{equation}
    \label{N1}
    N_1=\left( \frac{\Ne \Nc}{Q \T A_0^* b^3} \right)^{1/4}
    \end{equation}
A second critical length, corresponding to $\A = \A_2$ and found 
in the same way, can also be computed: 
    \begin{equation}
    \label{N2}
    N_2 = \left( \frac{\Ne}{Q \T A_0^* b^3 \Nc} \right)^{1/2}
    \end{equation}

A schematic plot of the evolution of the adhesion energy 
throughout the different regimes is shown in Figure~4 (and will be 
discussed further in the next section).
    \begin{figure}
    \centering
    \includegraphics*[clip=true, bb=5.5cm 11.3cm 16cm 
    18.5cm]{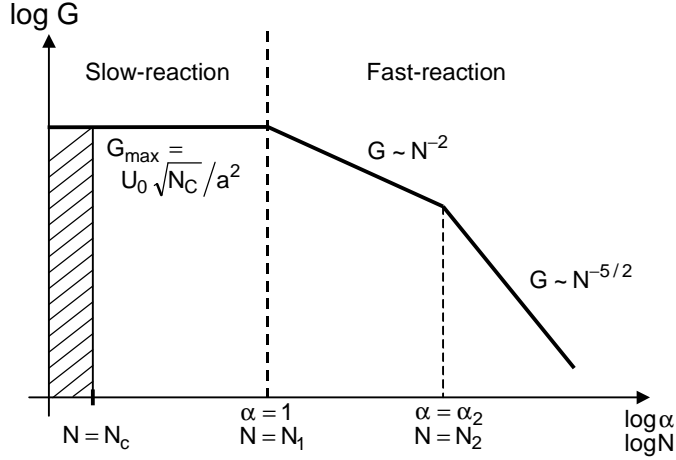} 
    \caption{Schematic plot of the adhesion energy $G$ throughout 
different regimes (see text), as a function of the control 
parameter $\A$ or of the chain length $N$ (in situations where it 
is the only free parameter). Note that values $N < \Nc$ are not 
considered as they do not result in the formation of a macroscopic 
network (gel).} 
    \end{figure} 
\section{Discussion}
\subsection{Comments on the results of the model}
In the previous section, we found several regimes for the final 
state of the interface: the slow-reaction regime, and the 
fast-reaction regime (with, for the latter, two sub-regimes). We 
transit from one regime to another by changing the value of $\A$ 
in the system, that is to say by changing the respective rates of 
the interdiffusion and the cross-linking reaction. A special case 
that may have practical applications is one where all the 
parameters characterizing the system are fixed, except for the 
chain length $N$. The transitions then occur for two critical 
values $N_1$ and $N_2$. 

The results of the computations of the preceding section can be 
summarized as follows (see Figure~4). For small $\A$ ($\A < 1$) or 
short chains ($N < N_1$), the interface is granted enough time to 
reach equilibrium before the reaction occurs. The interfacial 
strength is in this case maximum, with $G_{\ab{max}} \simeq U_0 
\sqrt{\Nc}/a^2$, and is equal to the (ideal) bulk strength. When 
$\A$ or $N$ is increased ($1<\A < \A_2,\, N_1 < N < N_2$), we 
enter a regime (fast-reaction regime) where the reaction occurs 
earlier, and does not allow for full equilibration: the 
interfacial energy starts decreasing (eq.~\ref{Gfastscission}) and 
displays an inverse dependence on $\A$ and $N$ of the form $G \sim 
\A^{-1/2} \sim N^{-2}$. Finally, if we increase $\A$ or $N$ 
further ($\A > \A_2, \, N>N_2$), the drop in the adhesion becomes 
even more pronounced (eq.~\ref{Gfastextraction}): $G \sim N^{1/2} 
\A^{-3/4} \sim N^{-5/2}$. 

On a practical standpoint, our conclusion is that for an optimized 
adhesion, i.e., optimized interface-related properties, we need 
(as expected) to place the system into the slow-reaction regime by 
a proper choice of the essential parameters involved in $\A$ (or 
by a proper choice of $N$ if $N$ is the only adjustable 
parameter). If, for some reason, such a proper set of values of 
the parameters is not accessible experimentally (for example, 
because the cross-link concentration has been chosen very high, 
and $\Nc$ is thus very small), we have predictions for the loss in 
interface energy that should be expected (fast-reaction regime). 

At this point, it might be appropriate to give a numerical 
example. A typical value for the critical chain length $N_1$ 
(below which we want to operate) with $\Ne = 100$, $\Nc = 500$, $Q 
\T = 10^{-9}$, $A_0^* b^3 \simeq A_0^* a^3 = 0.1$ (i.e.\ one A* 
site every ten polymer units) is found to be (eq.~\ref{N1}): $N_1 
= 4700$ units (a reasonable number). In this case, when $N < N_1$, 
with $U_0 = 300 \,$kJ/mol $= 3.1 \,$eV per monomer and $a = 5 
\,$\AA, the adhesion energy amounts to $G_{\ab{max}}=45 \, 
\mbox{J/m}^2$. 

This value for the adhesion energy is not a very large one, but it 
must be kept in mind that this is really the lower-bound of 
observable energies, since it corresponds to a zero-velocity 
fracture propagation. In practice, much higher values are reached 
--- hopefully, our estimations provide nevertheless useful 
guidelines toward the most advantageous situations.

\subsection{The marginal case $\Nc \simeq N$}
We have worked until now with the assumption that the cross-link 
density, and, accordingly, $\Nc$, are determined by the 
requirements on the bulk properties, and we optimized the 
interface strength under this constraint. What happens if the 
constraint can be relieved, and we are allowed to choose $\Nc$ 
freely? Suppose that we still want to enhance the interfacial 
properties: once we have placed ourselves into the slow-reaction 
regime, with an adhesion energy $G \simeq U_0 \sqrt{\Nc}/a^2$, it 
is easily seen that $G$ will increase only if we increase $\Nc$, 
i.e. have a looser network with less cross-links. If we really do 
not care much about the consequences on the bulk strength, we can 
reduce the cross-link density to the lowest possible value, i.e. 
two crosslinks per chain ($\Nc = 0.5 N$): we shall call this limit 
where $N \simeq \Nc$, the ``marginal case''. (This marginal case 
was the one we studied in ref.~\citen{Aradian}.) In the 
slow-reaction regime, when $N \simeq \Nc$, the adhesion energy 
(see eq.~\ref{Gslow}) becomes $G \simeq U_0 N^{1/2}/a^2$, and is 
now an increasing function of the chain length $N$. However, if 
the chain length is increased too much, we fall into the 
fast-reaction regime again (since $\A$ increases too). There is, 
however, a minor change as compared to the ``usual'' fast-reaction 
regime described in the previous section: the critical chain 
lengths $N_1$ and $N_2$ here collapse one onto the other, so that 
the first sub-regime (scission) disappears, and we enter directly 
into the second (extraction-dominated) sub-regime, which can be 
shown~\cite{Aradian} to behave like $G \sim N^{-7/4}$. 

The whole evolution of $G$ vs.\ chain length in this marginal 
regime is summarized in Figure~5. As can be seen, there is an 
optimum in the curve at the transition between the slow- and 
fast-regime. In our approach, this optimum energy is predicted to 
be the maximum attainable adhesion energy in the system (when all 
constraints on the bulk properties are released). The optimum 
chain length $N_{\ab{opt}}$ is found by letting $\A = 1$, 
    \begin{equation}
    \label{Nopt}
    N_{\ab{opt}} \simeq \left( \frac{\Ne}{Q \T A_0^* b^3} \right)^{1/3}
    \end{equation}
and the optimum energy is then simply 
    \begin{equation}
    \label{Gopt}
    G_{\ab{opt}} \simeq U_0 N_{\ab{opt}}^{1/2}/a^2 
    \end{equation}
    \begin{figure}
    \centering
    \includegraphics*[scale=1.3, clip=true, bb=6.5cm 12.8cm 14.5cm 
    17.5cm]{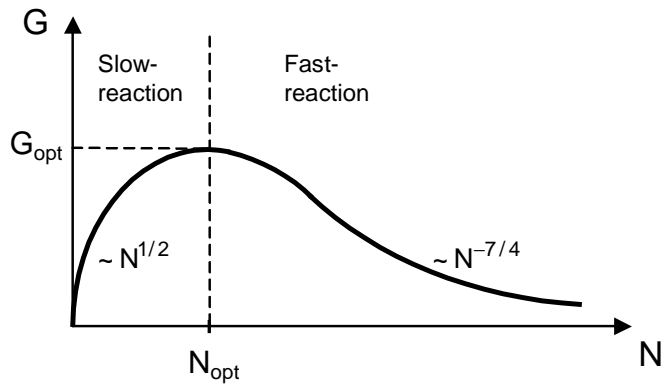} 
    \caption{Schematic plot of the adhesion energy $G$ vs.\ the chain 
length $N$ in the marginal case $\Nc \simeq N$.} 
    \end{figure} 
We should caution, however, that the formulae~\ref{Nopt} 
and~\ref{Gopt} are based on extrapolations near $\A = 1$ of the 
results found at $\A \ll 1$ or $\A \gg 1$, and should therefore be 
regarded as tentative. 

We may again give a numerical example. For $\Ne = 100$, $Q \T = 
10^{-9}$, and $A_0^* b^3 = 0.1$, we have $N_{\ab{opt}} = 10000$ 
units. With $U_0 = 300 \,$kJ/mol $= 3.1 \,$eV per monomer and $a = 
5 \,$\AA, the corresponding energy is $G_{\ab{opt}}=200 \, 
\mbox{J/m}^2$. 

We close this section by emphasizing that this ``marginal'' regime 
is not very realistic for latex coatings, where a high level of 
cross-linking is usually desired. But it may be applicable to 
certain types of low cross-link adhesives.
\subsection{Concluding remarks and further perspectives}
The approach to the competition of interdiffusion and 
cross-linking at interfaces presented in this article remains 
clearly very simplified, and we certainly do not hope to provide a 
fully quantitative description of the complexity inherent to real 
systems. Rather, the goal of this work was to try to extract the 
essential parameters of such systems, and see how they combine 
together in governing the dynamics (inside the parameter $\A$) or 
in determining the final state of the interface (inside the 
adhesion energy $G$).

We should emphasize that the value of the control parameter $\A 
\simeq Q \T A_0^* b^3 N^4/(\Ne \Nc)$ can in pratice be modified by 
many means: in section~4, we focused on the role of the chain 
length $N$, but this might not be the most relevant procedure from 
an industrial point of view. One could also change the ratio 
between ``active'' A*-sites and ``passive'' A-sites along polymer 
chains (thus affecting the concentration $A_0^*$), or change 
temperature (which would affect both $Q$ and $\T$). Indeed, in a 
recent experimental work on latex films,~\cite{LiuWinnik} Liu et 
al. studied (using fluorescence techniques) how the extent of 
coalescence between neighboring particles varied when the relative 
rates of interdiffusion and reaction were changed, either by way 
of temperature, or by adding an acid-catalyst that promoted the 
cross-linking reaction (i.e., in our language, increased $Q$). It 
is reasonable to think that this kind of systems would probably 
provide a direct and powerful approach to test the validity of our 
model. 

To conclude this article, we wish to present now a list of some
remaining issues, which, in our view, would have to be understood 
and included into a comprehensive theory. 

To start with, let us recall that we restricted the study to the 
case of a symmetric interface with two identical polymers facing 
each other. There are however many other possibilities (we refer 
the reader to section~4.3 of ref.~\citen{Aradian} for the 
corresponding discussion). 

One uncertain point is related to materials with a very high 
degree of crosslinking, like many industrial latex coatings, where 
the distance between cross-links becomes significantly smaller 
than the distance between entanglements ($\Nc \ll \Ne$): as of 
now, it is not clear to us whether our approach can be safely 
extrapolated to such situations, and if not, how it should be 
amended. 

A major issue is also to understand all these situations which 
have as a common feature to display \emph{inhomogeneities}, as 
they are in fact ubiquitous in practice. For instance, in latex 
films, the stability of the initial colloidal dispersion is 
ensured by coating the individual polymer particles with a layer 
of a surfactant or a charged polymer. But these surface layers, at 
the time when the particles come into contact, may significantly 
affect the interdiffusion dynamics.~\cite{Joanicot} A closely 
related difficulty arises when the polymer particles themselves 
are structured, with a core and a shell that often have very 
different properties (for example a different $T_{\ab{g}}$). 
Another example where inhomogeneities are crucial is found in the 
formation of a macroscopic joint between two elastomers: a 
frequent complication is the formation of an ``interphase'', i.e.\ 
a region (sometimes fairly broad) around the original interface, 
where the properties of the material present significant 
gradients.~\cite{Vallatinterphase} Quite often, this feature 
originates in a non-uniform distribution of the cross-linking 
agent in the initial situation (for example, more abundant at one 
side of the interface), which then diffuses (with non-uniform 
concentration) towards the interface. All the interfacial 
evolution and the final properties become then quite different. 

Another unsolved problem arises when the cross-links are not 
permanent, and have a certain ability to unfasten over long 
timescales. They may then migrate inside the material and modify 
the properties of the material in the bulk and at the interfaces. 

Chain length polydispersity should also play a role in the 
competition between interdiffusion and cross-linking. There might 
be interesting effects, due to the rapid diffusion of smaller 
chains towards the interface: these might form a ``crust'' upon 
cross-linking and considerably modify later diffusion of higher 
molecular weights fractions.\cite{Vallatprivate} However, at 
present, we still lack experiments on carefully controlled systems 
(for example, with bimodal samples containing a mixture of small 
and long molecules) to draw sound conclusions. 

Finally, an interesting line of thought for future work would be 
to explore analogies (and possible transpositions) of the approach 
presented here to situations where \emph{phase separation} 
--- rather than coalescence --- competes with cross-linking, as in 
thermoset-thermoplastic blends.~\cite{Referee} These materials are 
obtained from an initially homogeneous mixture of the 
thermoplastic polymer with the thermoset precursor. The latter is 
submitted to a cure, and as the reaction proceeds, the mixture 
usually destabilizes: a phase separation starts, and enters into 
competition with the cross-linking reaction, which reduces chain 
mobility through branching processes. We hope to investigate this 
aspect in the future. 
\section*{Acknowledgements}
The authors are indebted to M.\ Winnik for initiating this work 
and for many subsequent written and oral exchanges. They would 
also like to thank C.\ Creton, L.\ Léger and M.-F.\ Vallat for 
stimulating discussions and comments. 
\section*{Appendix. Computation of the connector density 
$\ts$}
In this appendix, we present an estimation of the density of 
connectors per unit area of interface $\ts$, which was used to 
estimate the adhesion energy given in eq.~\ref{Gfastscission} for 
the fast-reaction regime (more precisely, in the 
scission-dominated sub-regime delimited by $1<\A<\A_2$, or, 
equivalently, by $N_1<N<N_2$). 

In order to make a correct estimation of the connector density, we 
must consider in some detail the spatial conformation of the 
chains lying at the interface. Initially (i.e., at $t=0$), the 
interfacial chains are contorted because they are not allowed to 
cross the interface and have to ``reflect'' on it. Once the 
contact between neighbouring polymer pieces has been made, they 
start to relax towards their equilibrium Gaussian conformations 
(thereby crossing the boundary). However, in the fast-reaction 
regime under consideration here, these chains have time to relax 
only partially: only the chain portions that could escape from the 
initial, ``reflected'' reptation tube, are finally really able to 
cross the interface. In average, these relaxed portions have a 
contour length given by $S_{\ab{final}}$ (eq.~\ref{Sfinal}), and 
contain a number $p$ of monomers $p \simeq S_{\ab{final}}/a \simeq 
L_{\ab{final}}^2/a^2$. Moreover, because of their random shape, 
each of the relaxed portion crosses the interface \emph{several} 
times (and not just once as represented, for the sake of pictorial 
clarity, on Figures~2 and~3). As a consequence, when the reaction 
occurs and divides all chains into segments of $\Nc$ monomers, a 
given chain near the interface can, through its relaxed portions, 
find itself with \emph{several} such segments bridging the 
interface. In other words, a given chain may contribute, after 
reaction, to several connectors for the adhesion, and this is why 
the density $\sigma$ of chains crossing the interface 
(eq.~\ref{crossingdensity}) differs from the density of connectors 
$\ts$ which is involved in the Lake and Thomas formula for the 
adhesion energy $G$ (eq.~\ref{scission}). 

We thus have to find out which relation holds between $\sigma$ and 
$\ts$, and, for that purpose, the question that must be answered 
is the following: considering a relaxed chain portion (with $p 
\simeq L_{\ab{final}}^2/a^2$ monomers) which crosses the 
interface, how many connectors of $\tn \simeq \Nc$ monomers does 
it provide after reaction? 

The most convenient method is to divide the polymer chains into 
``blobs'', which each correspond to a chain segment of $\Nc$ 
monomers. Such blobs have a (Gaussian) radius $\Rb = a \Nc^{1/2}$. 
We may now think on the scale of blobs: all relaxed chain 
portions, with $p$ monomers, are seen as made with $p/\Nc$ blobs. 
The idea to find the number of connectors is then simple: we look 
at the spatial conformation of each chain portion (of blobs), and 
count how many blobs intersect the interface. Each of these blobs 
bridges the interface, and, accordingly, must count for one 
connector. Now, counting the number of such intersections is 
possible at the scaling law level: it can be shown that if we cut 
through a random walk of $q$ links with a virtual plane (passing 
through the origin of the walk), that plane is intersected an 
average of $q^{1/2}$ times. In our case, each relaxed portion of 
interfacial chain contains $p/\Nc$ blobs, and therefore intersects 
the interface $(p/\Nc)^{1/2}$ times. Thus, each interfacial chain, 
through its relaxed portion, provides $(p/\Nc)^{1/2}$ connectors. 

The density of \emph{connectors} $\ts$ is then simply obtained 
from the interfacial chain density as $\ts = \sigma \times 
(p/\Nc)^{1/2}$. Substituting with the value $p \simeq 
L_{\ab{final}}^2/a^2$, we finally find that the density of 
connectors $\ts$ is given by $\ts \simeq \sigma \times 
L_{\ab{final}}/(a \sqrt{\Nc})$, which was precisely the formula 
used in the main text to establish eq.~\ref{Gfastscission}. 

\begin{thebibliography}{99}
\bibitem{chainendsdistri}
Assuming a uniform distribution of chain ends in the material 
amounts to assume that there is no particular attraction or 
repulsion of chain ends towards the interface in the initial 
state, which may prove incorrect in certain situations.
\bibitem{density}
The initial volume density $\rho_0$ in free chains is reduced when 
the initial gel fraction in the polymer material is increased. If 
the initial gel fraction is zero (all the chains are linear and 
free), $\rho_0$ becomes maximum: $\rho_0 \simeq 1/(N a^3)$. 
Usually, the gel fraction inside the initial particles is 
non-zero, but not too high either (a significant amount of free 
chains must remain to yield good interdiffusion), so that we 
assume that this diminution of the density of free chains is 
properly accounted for by placing a purely numerical prefactor in 
front of that maximum value, without changing the scaling law.
\bibitem{Referee}
This possible analogy was brought to the attention of the authors
by one of the referees.
\end{thebibliography}
\end{document}